# Electron Weibel instability induced magnetic fields in optical-field ionized plasmas

Chaojie Zhang[1], Yipeng Wu[1], Mitchell Sinclair[1], Audrey Farrell[1], Ken A. Marsh[1], Jianfei Hua[2], Irina Petrushina[3], Navid Vafaei-Najafabadi[3], Rotem Kupfer[4], Karl Kusche[4], Mikhail Fedurin[4], Igor Pogorelsky[4], Mikhail Polyanskiy[4], Chen-Kang Huang[5], Wei Lu[2], Warren B. Mori[1,6] and Chan Joshi[1]

[1] *Department of Electrical and Computer Engineering, University of California Los Angeles, Los Angeles, CA 90095, USA*

[2] *Department of Engineering Physics, Tsinghua University, Beijing 100084, China*

[3] *Department of Physics and Astronomy, Stony Brook University, New York, NY 11794, USA*

[4] *Accelerator Test Facility, Brookhaven National Laboratory, Upton, NY 11973, USA*

[5] *Institute of Atomic and Molecular Sciences, Academia Sinica, Taipei 10617, Taiwan*

[6] *Department of Physics and Astronomy, University of California Los Angeles, Los Angeles, CA 90095, USA*

## Abstract

Generation and amplification of magnetic fields in plasmas is a long-standing topic that is of great interest to both plasma and space physics. The electron Weibel instability is a well-known mechanism responsible for self-generating magnetic fields in plasmas with temperature anisotropy and has been extensively investigated in both theory and simulations, yet experimental verification of this instability has been challenging. Recently, we demonstrated a new experimental platform that enables the controlled initialization of highly nonthermal and/or anisotropic plasma electron velocity distributions via optical-field ionization. Using an external electron probe bunch from a linear accelerator, the onset, saturation and decay of the self-generated magnetic fields due to electron Weibel instability were measured for the first time to our knowledge. In this paper, we will first present experimental results on time-resolved measurements of the Weibel magnetic fields in non-relativistic plasmas produced by Ti:Sapphire laser pulses (0.8 μm) and then discuss the feasibility of extending the study to quasi-relativistic regime by using intense $CO_2$ (e.g., 9.2 μm) lasers to produce much hotter plasmas.

## Introduction

The mechanisms of magnetic field generation commonly seen in terrestrial, space and cosmic plasmas have been a long-standing enigma in plasma physics. One well-known mechanism of self-seeding and amplifying magnetic fields in plasmas is the Weibel instability, first proposed by E. S. Weibel in 1959[1]. In the original formulation, this instability is driven by the self-organization of microscopic currents in stationary but anisotropic plasmas. Here anisotropic means the plasma has different electron temperatures (or electron velocity distributions, EVDs) along different spatial directions. In Weibel's original work, the plasma had a bi-Maxwellian EVD. As such a plasma approaches the thermal equilibrium, the attraction (repulsion) of co- (counter-) propagating plasma currents generates magnetic fields that receive energy from the kinetic energy of electrons. As the instability grows and both magnetic fields and collisions cause electron trajectories to bend, the plasma progressively isotropizes. During this process, the strength, wavevector spectrum and topology of the magnetic field evolve as a result of the continuous merging of plasma currents[2,3]. Fried later explained the onset of the Weibel instability using the electromagnetic two-stream picture, often referred to as the "current filamentation instability" (CFI)[4].

The Weibel/current filamentation instability has attracted a renewed interest in recent year due to its potential importance in laboratory and astrophysical plasmas- see ref 5 and references therein. For instance, Weibel magnetic fields are thought to be a seed for turbulence and dynamo amplification in galactic plasmas[6]. Weibel instability is also thought to play important roles in many scenarios that involving matter at extreme conditions, such as gamma-ray bursts (GRBs)[7,8], relativistic jets in active galactic nuclei (AGNs)[9], collisionless shocks[10,11], inertial confinement fusion[12–14], as well as electron-positron[15,16] and quark-gluon plasmas[17].

## Existing experimental approaches and our new platform

Because of its broad relevance, Weibel instability has been extensively studied in both theory and particle-in-cell (PIC) simulations. However, the experimental study of Weibel instability has proven challenging as it requires both the creation of anisotropic plasmas and high-resolution spatiotemporal measurements of fast-evolving magnetic fields. In the past decade several experimental approaches have been established, most of which are particularly suitable for studying CFI, of either ions or electrons. In Fig. 1(a)-(c) we show three typical experimental approaches.

The method shown in Fig. 1(a) uses multiple laser pulses with total energy from kJ to MJ to generate ablation plasmas by blasting two solid targets (e.g., CH foils) arranged face-to-face and separated by between a few millimeters to centimeters. These plasmas then expand and collide with each other to trigger the growth of ion CFI, since the energy in these flows is predominantly carried by ions. The growth of magnetic fields is probed by proton bunches generated by a separate synchronized laser pulse (laser-driven proton radiography)[18]. Typical parameters for such experiments can be found in references 19–22. The characteristic filamentary structure of CFI magnetic fields and the Bierman battery effect have been identified using this platform[19,21]. More recently, a modification of this platform in which the two solid targets were tilted to generate plasma flows that collided at a 130° angle was used alongside Thomson scattering of an external optical probe to record the evolution of current filaments moving through a fixed scattering volume[23]. This approach and its variations have proven quite effective in probing high energy density plasmas. A major obstacle preventing the widespread use of this approach, however, is that it requires energetic lasers that are only available at large facilities and typically operate at very low repetition rates (on the order of a few shots per hour or day).

The second approach as illustrated in Fig. 1(b) is suitable for investigating relativistic electron CFI. In this approach, an electron bunch from either a linear or laser wakefield accelerator propagates through a stationary plasma (either underdense[24,25] or overdense[26–28]). The electron bunch is modulated and eventually breaks into filaments due to electron CFI. In the underdense case, the optical transition radiation (OTR) generated when the filamented beam passes through a metallic foil is used to study the filamentary structure of the bunch after its interaction with the plasma. In the overdense case, an ultra-relativistic electron bunch is focused onto a metallic foil, and the intense magnetic fields generated by the electron CFI can bend the trajectories of beam electrons to emit bright gamma rays[26,29].

The third approach, sketched in Fig. 1(c), is used to study electron CFI in the quasi-relativistic regime. An intense laser pulse is focused onto a solid target to drive hot electrons with energy from a few tens to hundreds of keV on the surface of the target. As these hot electrons propagate forward into the target, cold electrons inside the target form backward propagating return currents, triggering the CFI. The magnetic fields growing in the surface layer of the target can then be measured using optical polarimetry[30–32].

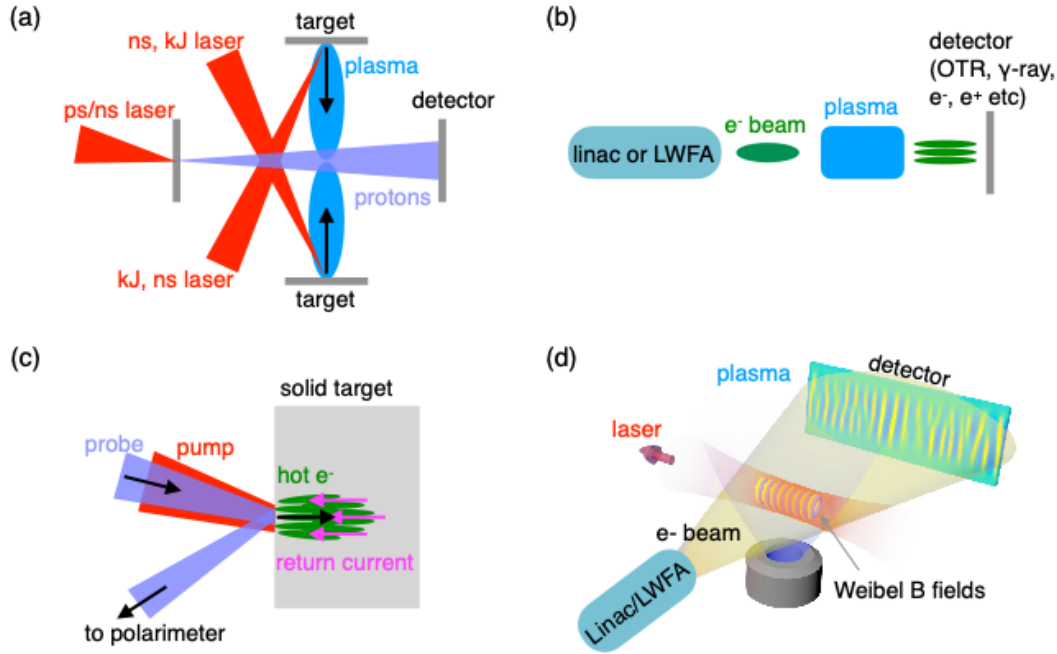

Fig. 1. Sketches of existing experimental approaches and our new platform. (a) Colliding plasma approach for probing ion CFI using laser-driven proton radiography or Thomson scattering. (b) Relativistic beam-plasma interaction for studying relativistic electron CFI. (c) Laser-solid interaction for investigating non- to quasi-relativistic electron CFI. (d) Our new platform that enables the investigation of thermal electron Weibel instability.

In all three approaches, the dominant mechanism for magnetic field generation is CFI driven by either ions or electrons. Experimental verification of thermal Weibel instability (the original concept of an electron Weibel instability driven by a temperature anisotropy in a stationary plasma [33]) has proven elusive until recently [34]. The main challenge is the lack of a suitable platform for controlled initialization of anisotropic EVDs and tracking of the evolution of the instability magnetic fields with high spatiotemporal resolution. In this paper, we show such a platform. A sketch of our platform is shown in Fig. 1(d). Here the initialization of a stationary anisotropic plasma is done via optical-field ionization (OFI) and the magnetic fields growing in the plasma are recorded by measuring the deflections of high-quality relativistic electrons from a particle accelerator with μm and ps spatiotemporal resolution.

## Preparing anisotropic plasmas using optical field ionization

To investigate the thermal electron Weibel instability, the first step is to initialize anisotropic plasmas in a controllable manner. This is done using ultrafast optical-field ionization (also called tunnel ionization since the Keldysh parameter $\gamma < 1$) of neutral atoms/molecules using ultrashort but intense laser pulses. For electrons released inside the laser field, the conservation of canonical momentum implies that by the time the laser pulse is gone, the electrons end up with finite residual momentum, for instance, $p_\perp = -\frac{e}{c}\boldsymbol{A}_\perp(t_0)$ in the transverse plane, where $\boldsymbol{A}_\perp$ is the transverse component of the laser vector potential and $t_0$ is the instantaneous time when the ionization happens. The resulting EVD is predominantly determined by the ionization potential of the atom, wavelength and polarization of the laser. By changing these parameters, we have shown that it is possible to create plasmas that are far from thermal equilibrium and have very large anisotropies[35,36].

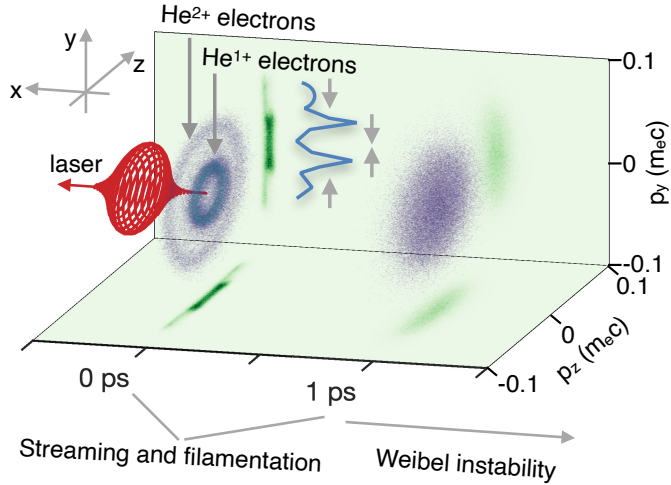

Fig. 2. Electron velocity distribution (EVD) of helium plasma ionized by a circularly polarized Ti:Sapphire laser. The EVD is extracted from a 3D PIC simulation which self-consistently model both the ionization process and the subsequent motion of free electrons in the laser field. The scatter plots of the electrons at two representative times (0 ps, right after the laser and 1 ps) as well as their projections along the y and z directions are shown. A hierarchy of the kinetic instabilities in this plasma is sketched (see text).

Figure 2 shows an example of the EVD of a helium plasma ionized by an ultrashort, circularly polarized (CP) Ti:Sapphire laser. The result is obtained from a three-dimensional (3D) PIC simulation using the code OSIRIS[37] where both the ionization and the subsequent motion of electrons are self-consistently modeled. The EVD in the plane of laser polarization at $t = 0$ (immediately after the passage of the laser) clearly shows two donut-shape structures corresponding to the two helium electrons with different ionization potentials. The reason for the EVDs having different radii for different ionization levels is that the two ionization potentials are reached by the laser fields at different field strengths. The plasma is significantly hotter in the plane perpendicular to the laser propagation direction, with a root-mean-square (rms) temperature of $T_\perp \approx 660$ eV. In the laser propagation direction ($x$), the plasma is cold ($T_\parallel \approx 5$ eV) because of the small component of the laser vector potential in this direction. This plasma thus has a very large initial temperature anisotropy $A = \frac{T_\perp}{T_\parallel} - 1 > 100$. We note that such an EVD has previously been measured in an experiment using Thomson scattering[35]. The EVD is not only highly anisotropic, but also contains interpenetrating streams along the radial direction. In such a plasma, there follows a hierarchy of kinetic instabilities that begin with the two-stream and current filamentation instability, which have also been measured using Thomson scattering of an external probe with fs resolution[36]. These instabilities reduce the temperature anisotropy very rapidly, from $A > 100$ to $A \sim 10$ in about one ps[36]. Figure 2 shows that the EVD of the plasma has relaxed to an approximately bi-Maxwellian (in each direction the EVD is nearly Maxwellian but with different temperatures) distribution at $t \approx 1$ ps, with $T_\perp \approx 500$ eV and $T_\parallel \approx 50$ eV. It is at this point that the Weibel instability becomes the dominant instability, which is the focus of the rest of this paper.

**Time-resolved measurements of the Weibel magnetic fields**

An experiment was performed to demonstrate the OFI induced Weibel instability platform sketched in Fig. 1(d). In this experiment, which was performed at Tsinghua University where a 10-TW Ti:Sapphire laser (0.8 μm) and a 50-MeV linear accelerator are collocated, ultrashort laser pulses were used to ionize a supersonic helium gas jet. The full width at half maxima (FWHM) of the laser pulse duration was about 50 fs. The laser was focused to a $w_0 \approx 22$ μm spot with a peak intensity of $\sim 2.5 \times 10^{17}$ W/cm[2] to rapidly ionize both of the helium electrons in just a few optical cycles through tunneling ionization. The laser intensity was kept low enough to prevent driving large-amplitude wakes. The electron bunch serving as the probe was delivered by a linear accelerator with a 3-m-long accelerating structure and the peak energy of the electron probe was 45 MeV (~0.5% energy spread). The probe bunch length was compressed down to $\tau_{FWHM} \approx 1.8$ ps with a total charge of ~30 pC. The perturbation of the plasma by the electron bunch was negligible due to the low beam current. Two quadrupole magnets were used to focus the bunch into an elliptical spot ($\sigma_{z,FWHM} \approx 1.2$ mm and $\sigma_{y,FWHM} \approx 0.4$ mm) at the interaction point (IP). The focal plane of the electron bunch was put upstream of the IP and therefore the beam was slightly divergent at the IP such that the geometric magnification at the detector plane was $M \approx$ 1.1x. The probe electrons were deflected by the force exerted by the quasi-static electromagnetic fields as they traversed the plasma. For the plasma density ($n_p \approx 10^{19}$ cm$^{-3}$) used in this experiment, the electric fields (e.g., wakes) inside the plasma which oscillate at the plasma frequency cannot be seen by the relatively long electron probe due to the averaging effect (the front and back parts of the probe see different phases thus the integration approaches zero)[38]. The magnetic fields, on the other hand, are quasi-static compared to the probe duration and the transit time. Therefore, the deflection of the electron probe was dominated by the magnetic fields through the $\boldsymbol{v} \times \boldsymbol{B}$ force. As the electron bunch further propagated in vacuum, the angular deflection of the probe electrons translated into density (flux) modulation which was captured by an electron imaging system consisting of a thin (100 μm) YAG:Ce crystal placed 23 cm away from the plasma followed by an optical relay system equipped with a CCD camera.

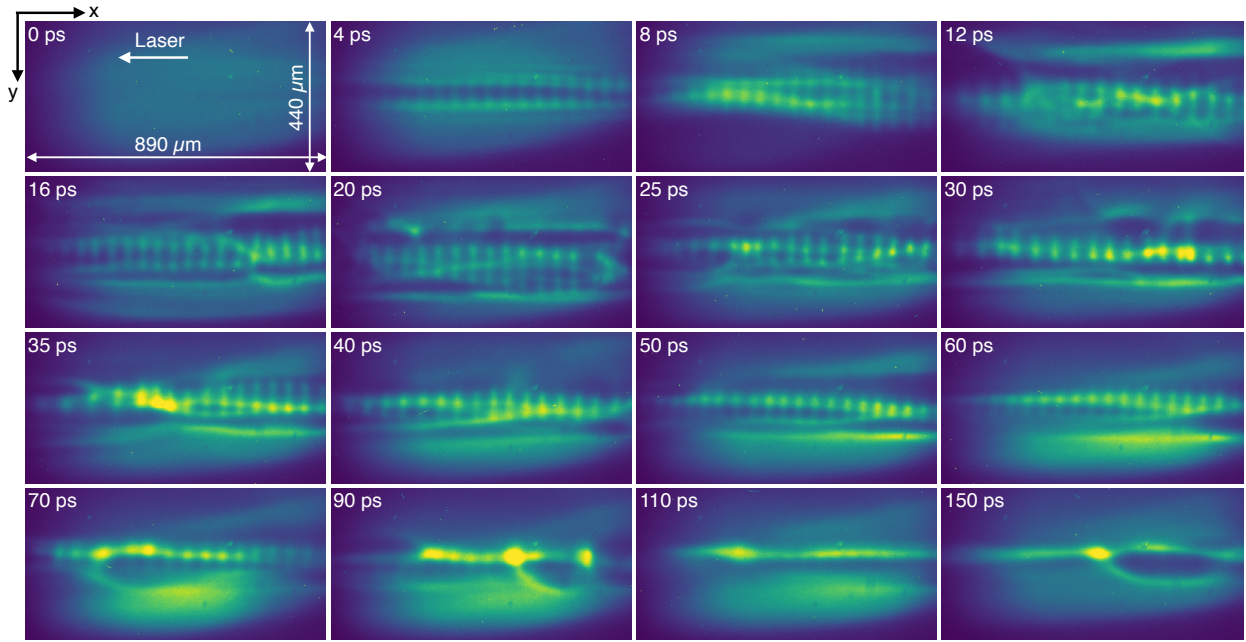

Fig. 3. Snapshots taken at different times of the flux modulations induced on the probe electron beam. The electron deflections are caused by the Weibel magnetic fields in helium plasma ionized by circularly polarized, 0.8-μm laser pulses. The laser propagates from right to left.

An example dataset is shown in Fig. 3. Each frame corresponds to a snapshot of the magnetic fields taken at a different delay with respect to the laser. The laser (not shown) propagates from right to left. Time zero is defined as the time when the electron beam overlaps with the laser at the IP, which is centered in the frame. The most noticeable features include vertical strips with regular spacing (short wavelength) and horizontal strips with longer wavelength. The modulation magnitude of the vertical strips (indicated by the brightness of the modulations) increases with delay, reaching the peak at ~20 ps, and then decreases. At later time (e.g., 90 and 150 ps), bright spots at the intersecting point of long-wavelength strips and dark regions surrounded by long-wavelength strips are visible. These features indicate the evolution of the magnetic fields in the plasma.

As we have explained in the previous section, the OFI plasma is hot in the laser polarization plane and cold in the other orthogonal direction. Theory shows that Weibel magnetic fields grow in such a way that the wavevector of the magnetic field is along the cold direction[1]. For this CP case, theory predicts that the Weibel magnetic fields should have a wavevector pointing along the laser propagation (horizontal) direction with a helicoid structure[33]. In other words, the expected probe density modulation should appear as vertical strips. Based on this justification, we can extract the contribution from the Weibel magnetic fields by isolating the short-wavelength structures using the method described in Ref. 34.

The probe electron density modulation is caused by the Weibel magnetic fields. For a given magnetic field wavelength $\lambda_B$ and parallel probe beam, the density modulation magnitude is proportional to the magnetic field strength as long as the normalized displacement (deflection) of the probe electrons $\mu \equiv \theta L / \lambda_B \ll 1$. Here $L$ is the drift distance from the plasma to the detector, and $\theta \approx \frac{e \int B dz}{\gamma mc}$ is the deflecting angle of the probe electrons induced by the magnetic field. The deflecting angle $\theta$ is proportional to the line integral of the $B$ field, $\int B dz$, and inversely proportional to the momentum of the probe electron. As $\mu$ becomes large (e.g., approaching unity), the mapping from the magnetic field distribution to the probe density modulation becomes highly nonlinear. For instance, the magnetic fields can be viewed as lenses for the probe electrons and caustics will form in the probe density profile if the detector is placed close to the focal plane of these lenses. In Ref. 39, these two cases are called the small- ($\mu \ll 1$) and large- ($\mu \gtrsim 1$) deflection regimes. In the small-deflection regime, the probe density modulation is proportional to the magnetic fields and therefore the temporal evolution of the magnetic fields can be well-approximated by that of the measured density modulation, from which we can extract the growth rate of the magnetic field[34].

**Retrieving magnetic fields**

More insights can be gained by retrieving the magnetic field distribution from the measured density modulation of the probe beam. The fields (or deflection angles) can be retrieved computationally by solving an equivalent optimal transport problem[40]. This approach assumes that the retrieved field should be distributed such that when it maps each small portion of the source profile to a new one with flux conserved, the overall transport of all these small portions

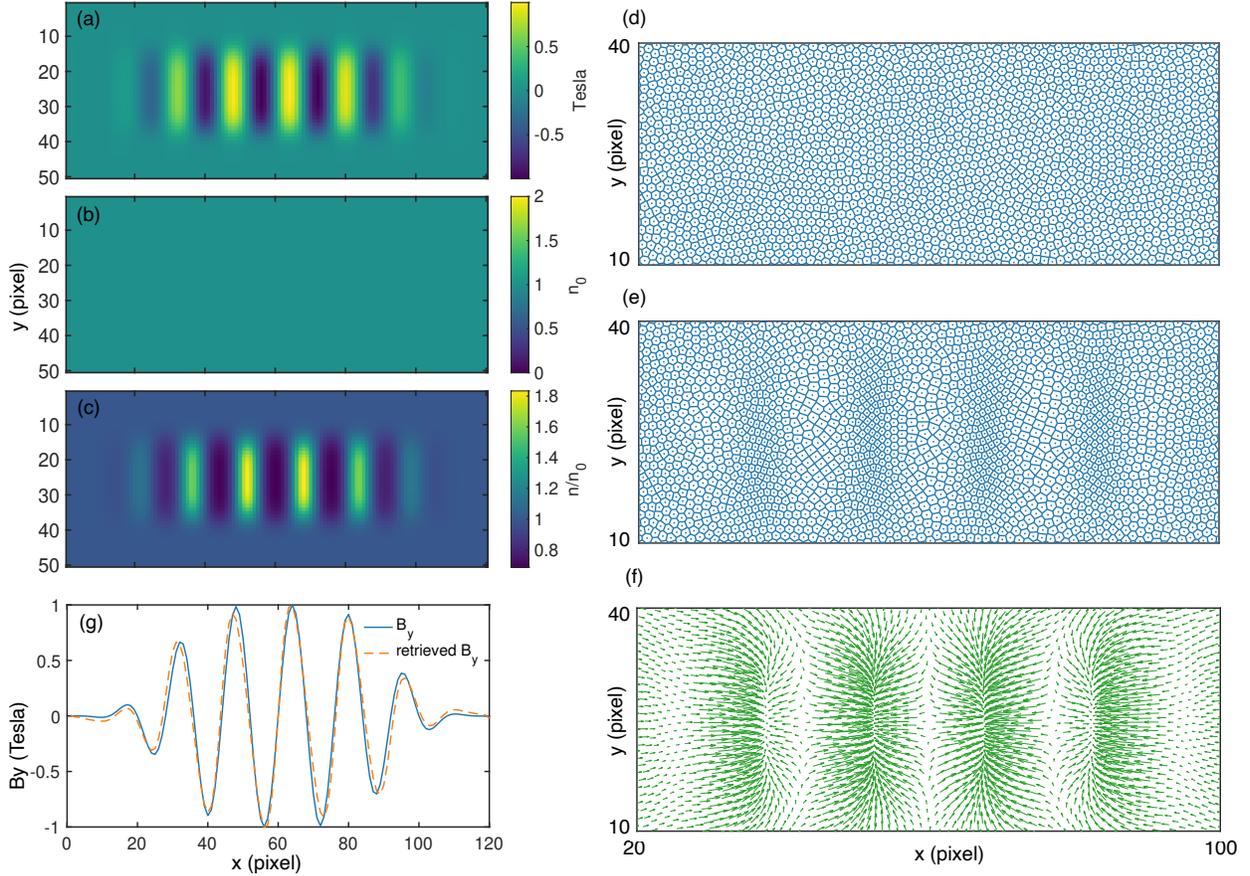

Fig. 4. Illustration of retrieving 2D deflection angles from probe beam density modulation by solving an optimal transport problem. (a) is a preset magnetic field $B_y$. (b) and (c) are the uniform unperturbed density (background) and modulated density (data) of the probe beam, before and after the beam traverses through the magnetic field. (d) is the initial Voronoi cells that sample the background and (e) is the Voronoi cells after optimization which maps out the density modulation in (c). The displacements of the probe beam are represented by the green arrows in (f). A comparison of the preset and retrieved magnetic field [i.e., axial lineout between (a) and (c)] is shown in (g).

An illustration of retrieving deflection angles from modulated probe density profile is shown in Fig. 4. A predefined magnetic field $B_y$ with peak magnitude of 1 Tesla and a wavelength of $\lambda_{Bx} = 160$ μm (the image pixel size is 10 μm) is shown in Fig. 4(a). The uniform background and a synthetic modulated density profile of the probe beam are shown in (b) and (c), respectively. Figure 4(d) shows a Voronoi tessellation[41] of the uniform background, with the flux in each cell being approximately the same. The weights of each cell are then iteratively modified until the power diagram properly maps cells from the source plane (b) to the measured density profile on the target plane shown in (c). Note that the flux in each weighted cell remains approximately the same. An example of the optimized distribution of the Voronoi cells is shown in Fig. 4(e). Now the displacement of each cell can be calculated and the result is shown in Fig. 4(f), where each arrow represents the vector displacement of one cell. The deflection angle along two orthogonal directions can then be retrieved given the probing geometry.

The deflection angles are related to the line integral of magnetic fields as follows,

$$\alpha_x \approx \frac{\Delta p_x}{p_z} = -\frac{q_e \int B_y dz}{\gamma m_e c}$$

$$\alpha_y \approx \frac{\Delta p_y}{p_z} = \frac{q_e \int B_x dz}{\gamma m_e c}$$

Using these equations, the line integral of the magnetic field components can be calculated. A comparison of the preset and retrieved magnetic field (axial lineout) is shown in Fig. 4(g). From this we can see that the retrieved magnetic field matches the predefined one fairly well.

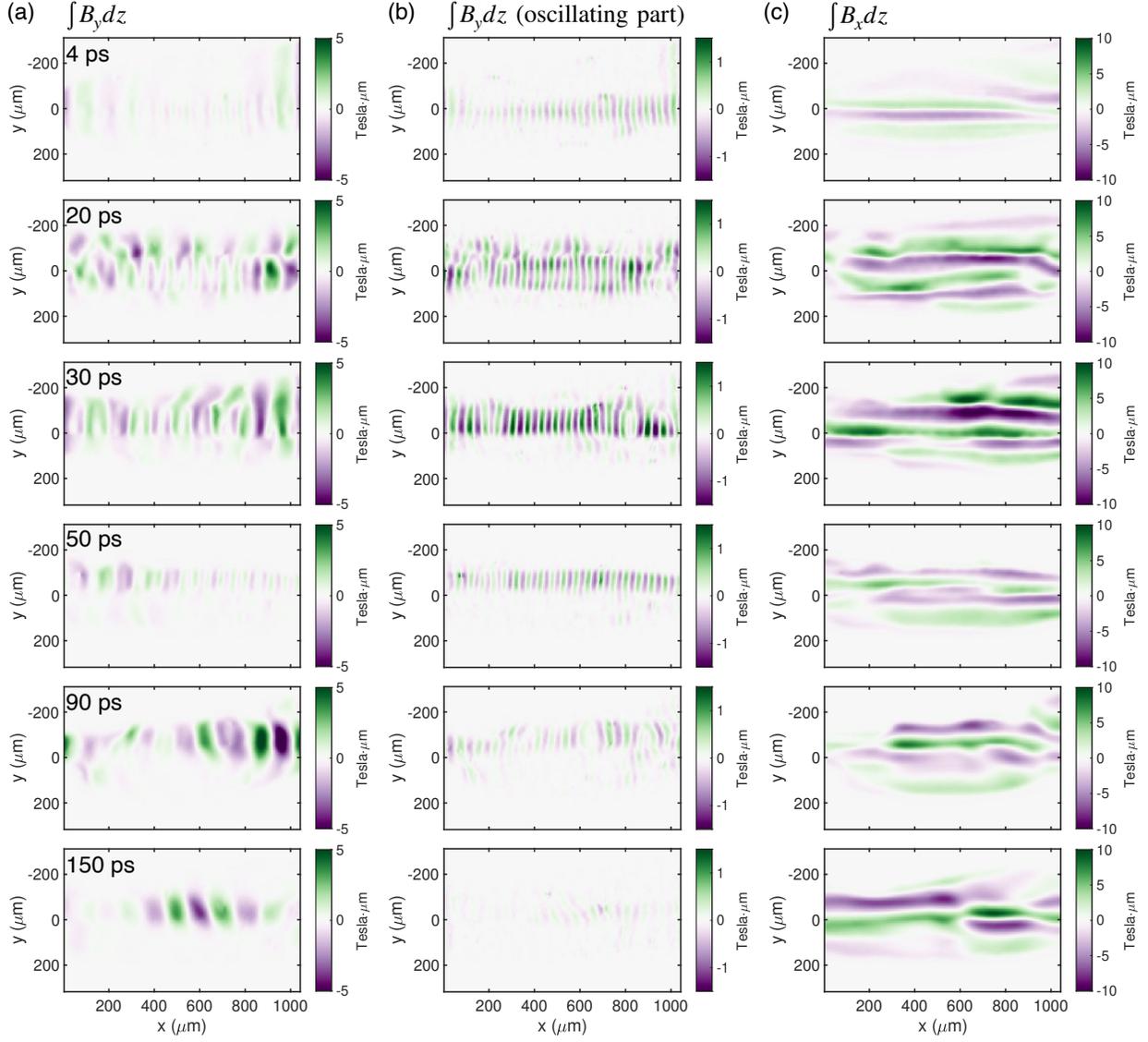

Fig. 5. Retrieved magnetic fields from the data in Fig. 3. Column (a) shows the retrieved line integral of $B_y$ along $z$. The oscillating part of (a) is shown in (b). The line integral of $B_x$ is shown in column (c).

Using the same procedure, we have retrieved magnetic fields corresponding to the snapshots in Fig. 3 and the results are shown in Fig. 5. Figure 5(a) shows the line integral of $B_y$ along $z$ (the probe direction). The oscillating part of $B_y$ is isolated by applying a high-pass filter, and the results are shown in (b). We note that the retrieval algorithm amplifies low-frequency (long-wavelength) components as mentioned in Ref. 40. This is why a high-pass filter is needed to isolate the short-wavelength Weibel fields that are of interest. The most noticeable feature of the retrieved $B_y$ field is that it has parallel vertical strips that are almost equally spaced and with alternating sign along the $x$ direction (laser propagation direction). This is consistent with the short-wavelength, vertical strips in the measured probe density profile shown in Fig. 3. The magnitude of the $B_y$ field in (b) initially increases with time until it reaches the maximum at about 25 ps, and then decays from there. This is also consistent with the evolution of the data shown in Fig. 3. The retrieved $B_x$ fields are shown in Fig. 5(c). In contrast to the $B_y$ field, the $B_x$ field shows as long, horizontal strips, which can bend the probe electrons along the $y$ direction to form the observed long-wavelength structure in Fig. 3.

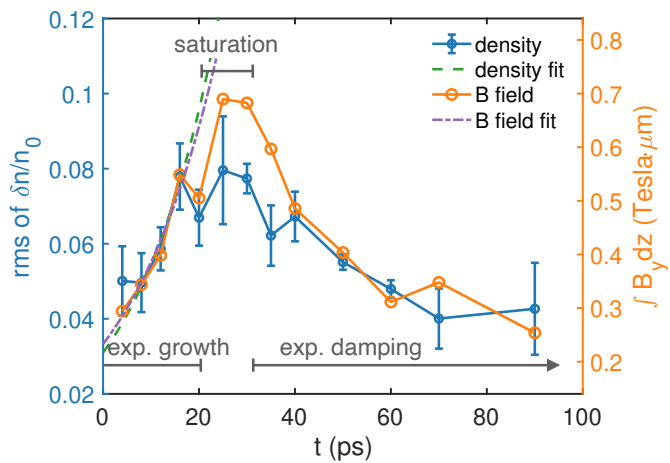

Fig. 6. Comparison of time evolution of probe density modulation (blue) and retrieved magnetic field strength (orange). The dashed and dash-dotted lines represent the best exponential fit to the density and field data, respectively.

As we have explained in previous sections, in the small-deflection regime, the modulation of the probe density profile gives a good representation of the actual magnetic field. This is illustrated in Fig. 6, where the root-mean-square of the measured probe density modulation ($\delta n/n_0$) as a function of time is plotted together with the evolution of the retrieved $B_y$ field (the oscillating part). The evolution of $B_y$ field tracks the probe density modulation very well except for when the field reaches the peak. This is due to the increased nonlinearity in the mapping between the field and the probe density profile as the magnetic field becomes strong for a given wavelength and probing geometry (in other words, $\mu$ increases beyond the small-deflection approximation). This suggests that the growth rate of the instability deduced using the probe density modulation is consistent with that extracted using the field data, with a relative difference of less than 10%. On the other hand, the saturation level should be evaluated using the retrieved magnetic field instead of the probe density modulation magnitude[34] to ensure accuracy.

**Linear polarization data**

One of the most attractive features of OFI plasma is that its initial EVD can be readily controlled by changing laser parameters such as polarization. Unlike in the circular case, a linearly polarized laser produces a plasma that is hot only along the laser polarization direction and colder in the other two orthogonal directions. Unlike in the CP case where the laser field only rotates so that when ionization happens the vector potential of the laser field is large, in the LP case, the laser field oscillate and when ionization happens at the peak of the electric field, the corresponding vector potential is small. As a result, the absolute plasma temperature in the LP case is also significantly lower. Both characteristics affect the growth of Weibel instability. In the experiment, we changed the laser polarization to linear (in the horizontal direction) and measured the evolution of magnetic fields. The results are shown in Fig. 7. For this dataset, the plasma density is the same as that of the dataset shown in Fig. 3.

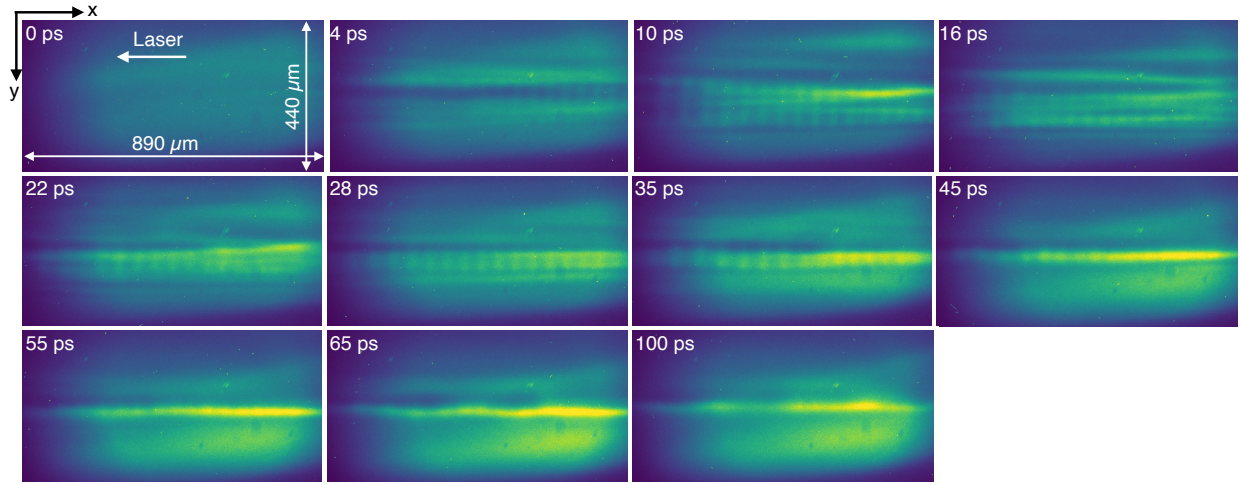

Fig. 7. Time-resolved measurements of the Weibel magnetic fields in helium plasmas ionized by linearly polarized, 0.8-μm laser pulses. The direction of polarization is in the $\hat{z}$ direction (in and out of the page). The laser is travelling to the right in $-\hat{x}$ direction.

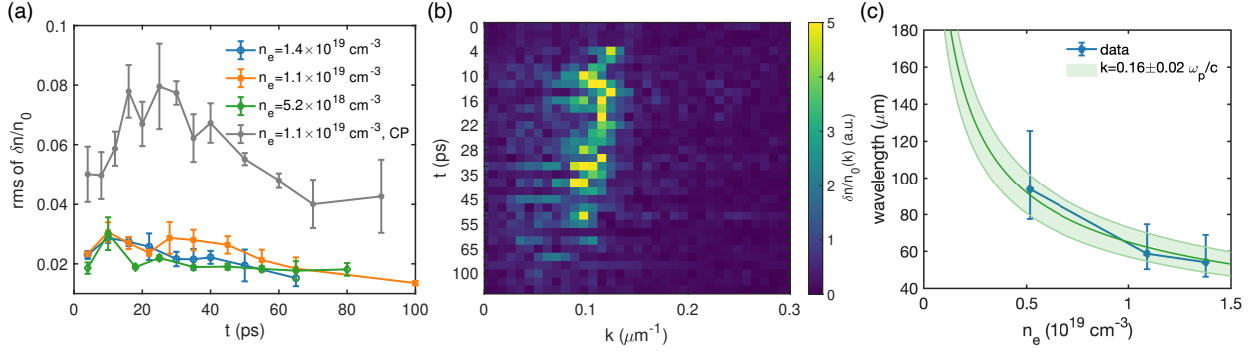

Fig. 8. Temporal evolution of the measured density modulation for the linear polarization case. (a) Density modulation as functions of time for three different densities. The grey line is a replicate of the circular polarization data shown in Fig. 6. (b) $k$ spectrum as a function of time for the $n_e = 1.1 \times 10^{19}$ cm$^{-3}$ case [orange line in (a)]. (c) Density dependence of the measured wavelength of the magnetic field.

The overall trend of the magnetic field evolution, shown in Fig. 7, is similar to that of the circularly polarized case (see Fig. 3)- once again we can see long-wavelength structures and superimposed short-wavelength vertical strips. There are also differences: the magnitude of the density modulation is smaller and disappears faster. In Fig. 8(a), we show the evolution of the probe density modulation for the data shown in Fig. 7, as well as at two other different densities. In all three cases, the signals reach saturation at ~10 ps compared against ~20 ps in the previous CP case. The apparent saturation level is also much smaller compared to the CP case (the grey line is a replicate of the CP data shown in Fig. 6). The approximately three times smaller density modulation magnitude upon saturation means that the magnetic field strength is also smaller for the same ratio since the wavelengths of the field are similar. This may be attributed to the lower plasma temperature which limits the amount of kinetic energy available to be converted to magnetic energy, and increases the rate of dissipation of magnetic fields due to collisions. The $k$ spectrum evolution for the $n_e = 1.1 \times 10^{19}$ cm$^{-3}$ case is shown in Fig. 8(b). Once again, a quasi-single mode is visible soon after the density modulation structure becomes observable. A slight shift of the spectrum towards smaller $k$ (longer wavelength) is visible which is consistent with the general trend predicted by theory. Figure 8(c) shows the wavelength of density modulation as a function of plasma density, where the green line shows the best fit to the data which gives $k \approx 0.16 \pm 0.02\ \omega_p/c$. This indicates that by the time we measured well-organized structures, the anisotropy of the plasma has dropped from an initially large value to a rather small one ($A < 1$), very similar to the CP case.

### Access to the quasi-relativistic regime using $CO_2$-driver

In the previous section, we have shown the possibility of controlling the initial EVD of an OFI plasma by switching the laser polarization and thus changing the evolution of Weibel magnetic fields. Another controllable experimental parameter is the laser wavelength. Using longer wavelength lasers can increase the plasma temperature. Since the residual momentum of an OFI electron is $p_\perp \approx -eA_\perp(t_i)/c = -a_0(t_i)m_ec$, where $a_0 \approx 8.5\lambda[\mu m]\left[I\left(10^{20}[\frac{W}{cm^2}]\right)\right]^{1/2}$ is the normalized vector potential of the laser. Compared to the 0.8 µm driver, switching to a $CO_2$ laser (e.g., $\lambda = 9.2$ µm), increases $a_0(t_i)$ by about a factor of 10 corresponding to a potential increase in plasma temperature by a factor of a hundred.

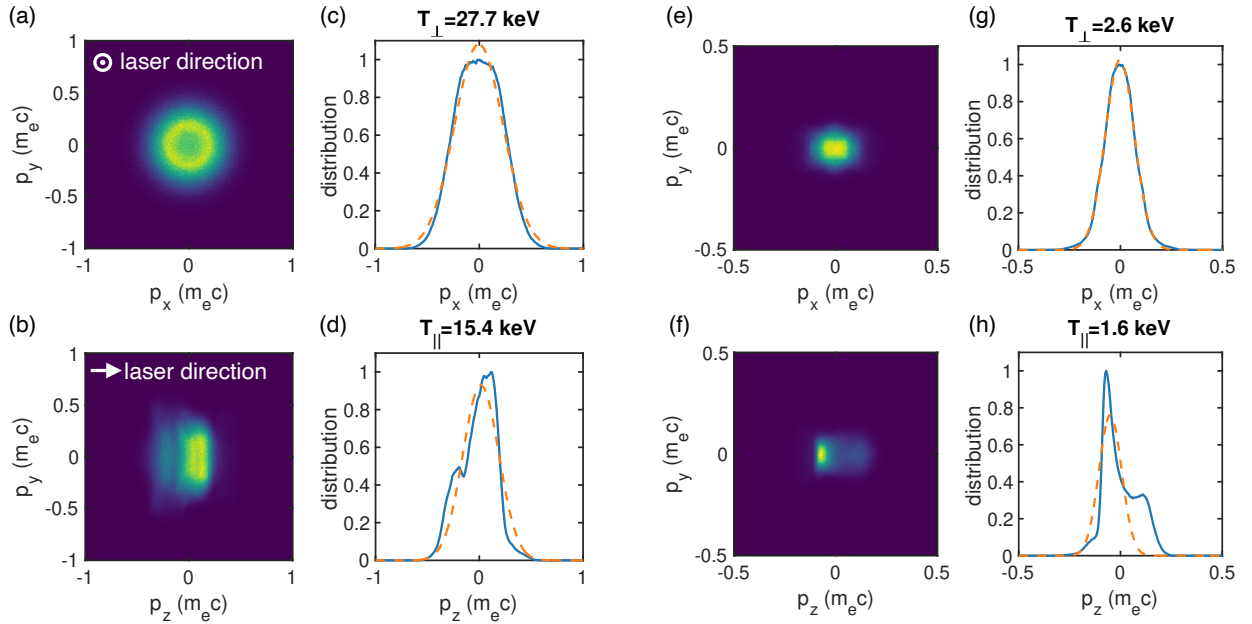

Fig. 9. Simulated EVDs of helium plasma produced by a 3-ps $CO_2$ laser with $w_0 = 50$ µm and plasma density of $n_p = 10^{17}$ cm$^{-3}$. (a)-(d) show the circular polarization results. (a) and (b) show the 2D velocity distribution in the transverse ($y$-$x$) and the longitudinal ($y$-$z$) plane, respectively. The projected distributions (blue solid line) and the corresponding Gaussian fits (dashed red lines) are shown in (c) and (d), respectively. The linear polarization results are shown in (e)-(h).

In Figure 9 we show two simulated examples of the EVDs of helium plasma ionized by a 3-ps, 10-μm $CO_2$ laser, with either circular [(a)-(d)] or linear [(e)-(h)] polarization. In the 3D PIC simulation, the laser is launched from the wall on the left side and focused onto the center of the helium gas initialized in the simulation box. The laser propagates along the $z$ direction. The vacuum focal spot of the laser is $w_0 = 50$ μm. The normalized vector potential is $a_0 \approx 0.86$ for both cases, which correspond to an intensity of $I = 2 \times 10^{16}$ W/cm$^2$ for the circular polarization case and a factor of two lower for the linear polarization case. The plasma density of the fully ionized region is $10^{17}$ cm$^{-3}$. Figure 9(a) plots the EVD of the helium plasma on the transverse $x$-$y$ plane, which is evaluated immediately after the laser passed. The EVD shows a ring structure as a result of the circular polarization. Compared to the 0.8-μm laser case (see Fig. 2) is that instead of two rings here only one is visible. This is due to the relatively long pulse duration of the laser that allows the EVD to significantly evolve by the time the laser is gone. Figure 9(b) shows the EVD in the $y$-$z$ plane. The integrated EVDs are shown by the blue lines in Fig. 9(c) and (d). In each plot, a gaussian fit is represented by dashed lines. The helium plasma ionized by a circularly polarized 0.8-μm laser has temperatures of $T_\perp \approx 0.5$ keV and $T_\parallel \approx 0.04$ keV. Here in the CP-$CO_2$ case, these temperatures increase to $T_\perp \approx 28$ keV and $T_\parallel \approx 15$ keV, which are in reasonable agreement with the $I\lambda^2$ scaling. The temperature anisotropy, however, is only $\approx 0.9$, much smaller than $\approx 10$ for the 0.8-μm case. This is a potential result of the finite bandwidth of the relatively long (but realistic) pulse length $CO_2$ due to collisional broadening effects, which may cause the plasma electron energy distribution to evolve significantly by the time the laser has passed. Simulations show that reducing the pulse length of the $CO_2$ laser to 0.3 ps gives plasma temperatures of $T_\perp \approx 47$ keV and $T_\parallel \approx 8$ keV. Figure 9(e)-(h) show the corresponding results for a linearly polarized $CO_2$ laser. As expected, the plasma is much colder compared to the CP case, but still much hotter than that produced by a 0.8-μm laser. The temperature anisotropy is also small ($\sim 0.6$) as in the CP case due to the relatively long pulse duration. These results demonstrate that by switching to a $CO_2$ laser driver, it is possible to significantly increase the plasma temperature to quasi-relativistic regimes, especially so with circular polarization.

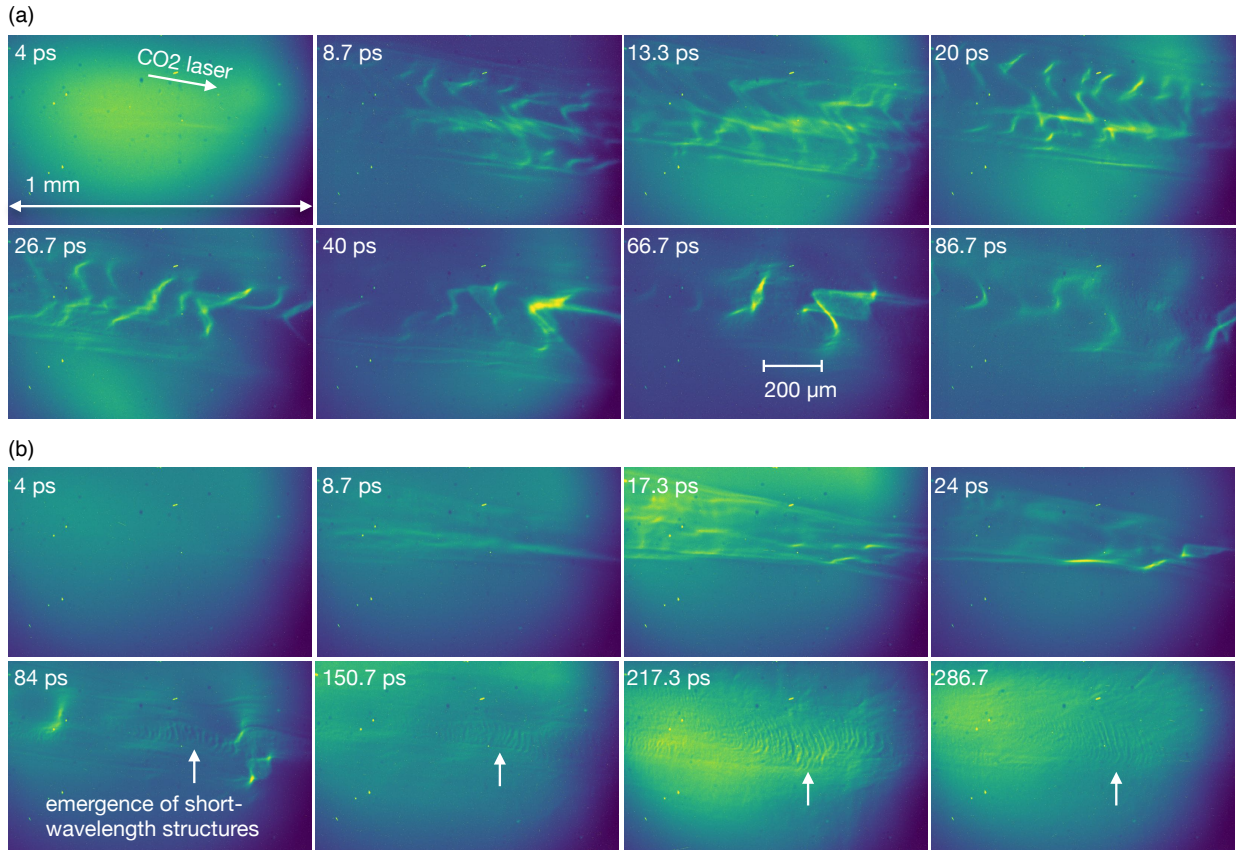

Fig. 10. Time sequences of snapshots of magnetic fields in $CO_2$ produced helium plasma. The laser is linearly polarized along the direction perpendicular to the screen. (a) High plasma density case, $n_p \approx 10^{18}$ cm$^{-3}$. (b) Low plasma density case, $n_p \approx 2 \times 10^{17}$ cm$^{-3}$. The white arrows mark the short-wavelength (~18 µm) structures emerged at later delay.

An experiment was performed at the Accelerator Test Facility (ATF) of Brookhaven National Laboratory (BNL) where a linear accelerator and a high-power $CO_2$ laser are co-located[42]. The experimental setup is similar to that in Fig. 1(d) except that a set of permanent quadrupole magnets (PMQs), with a designed beam energy of 50.5 MeV, were used to relay and magnify the electron image from an effective object plane that is just ~15 mm away from the plasma to the YAG:Ce screen ~0.5 m away from the plasma. The function of the PMQ set is to form a point-to-point image with a magnification of ~3.8x. The pulse length of the electron probe used in the experiment was ~1 ps. Because of the large spot size at the IP (~1 mm), the effect of the finite emittance of the probe beam on the final image was negligible. Time-resolved measurements of the beam deflections induced by the magnetic fields in the helium plasma produced by a 2-ps $CO_2$ laser (2-3 J energy, $I \approx 10^{16}$ W/cm$^2$) were taken and the snapshots of the fields are shown in Fig. 10(a) and (b) for two different plasma densities, $n_p \approx 10^{18}$ cm$^{-3}$ and $n_p \approx 2 \times 10^{17}$ cm$^{-3}$, respectively. In the higher density case, evolution of the structure from initially short-wavelength to later long-wavelength is visible. Similar to the results explained in previous sections, the approximately vertical strips are understood to be due to the Weibel magnetic fields, however in the experimental case these strips have more irregular shapes (somewhat twisted when compared with the straight strips in Fig. 3). This suggests that in this case the plasma may be more turbulent because the drive laser has a larger $a_0$ (~0.9 compared to ~0.1 in the 0.8-μm case) and therefore the ponderomotive force may be more strongly perturbing the plasma electrons. Another possible reason may be that at higher laser intensities the laser spot becomes more distorted and is better approximated by a sum of several Hermite-Gaussian modes of different amplitudes. A third reason may be the break-up of the laser pulse itself by the modulational instability that produces a relativistic electron plasma wave (often loosely called a wake). However, these wakes cannot be seen by the ps electron probe because of their fast-oscillating nature. Figure 10(b) shows the results of similar measurements taken with reduced plasma density. In this case, apart from the normal long-wavelength structures at early times (e.g., $\leq$24 ps), there is the emergence of intriguing structures with very short wavelength ($\lambda \approx 18$ μm) at larger delays (e.g., $\geq 84$ ps). These structures eventually disappear about 0.3 ns after the ionization of the plasma. The formation mechanism of these structures is presently not understood and will be studied further experimentally and with PIC simulations.

**Conclusion and outlook**

In summary, we have described a new experimental platform that is suitable for studying kinetic instabilities including streaming, current filamentation and Weibel instabilities in the laboratory. We have shown detailed measurements of the growth, saturation and damping of the thermal electron Weibel instability. We first show an example of the Weibel field evolution in helium plasmas ionized by circularly polarized 0.8-μm lasers. The density modulation of the probe beam induced by the quasi-static magnetic fields in the plasmas clearly shows the ultrafast magnitude and topology evolution of the magnetic fields. By solving an equivalent optimal transport problem, the 2D deflection angle distribution and thus the magnetic fields in the plane perpendicular to the probe direction were retrieved. The retrieved magnetic fields show that the magnetic fields along the laser propagation direction are the dominant components of the total fields. The Weibel magnetic fields grow on top of these long-wavelength structures and the data show that they grow and saturate in about 20 ps and then damp over the course of around 100 ps. In a previous publication, we extracted the growth rates of the Weibel instability directly using the measured density modulation data and that is justified here by comparing the evolution of the magnitudes of density modulation and the retrieved magnetic field. The effect of laser polarization and wavelength on the Weibel magnetic field evolution was investigated by switching the polarization from circular (800 nm laser) to linear ($CO_2$ laser). In the linear polarization case, the short-wavelength vertical strips are also visible but the absolute magnitude of the probe density modulation for both these short-wavelength and the horizontally aligned long-wavelength structures are smaller when compared against the circular polarization case. The saturated magnetic field strength in the linear polarization case is also several times smaller. The $k$ spectrum evolution and the dependence of field wavelength at saturation on plasma density are qualitatively similar to the circularly polarized case. By switching to longer wavelengths, we show in 3D PIC simulations that the plasma temperature can be significantly increased, demonstrating the possibility of accessing the quasi-relativistic regime.

The results shown in this paper highlight the great potential of this new paradigm where high-quality electron bunches from modern accelerators are used to explore transient electromagnetic fields in plasmas. There are several attractive advantages of the platform we described here compared to other complementary approaches such as laser-driven proton radiography. The high repetition rate (e.g., >1 Hz) of such beams allows the collection of large datasets and therefore enable quantitative statistical analysis. The stability and reproducibility of electron beams from linear accelerators are excellent, which significantly reduces the need to make assumptions of the background signal, making data analysis easier and more reliable. Additionally, the electron beam can be readily manipulated in terms of energy, charge, pulse duration, spot size, and divergence to provide versatile probing geometries.

**Acknowledgement**: This work was supported by the Office of Naval Research (ONR) Multidisciplinary University Research Initiatives (MURI) N00014-17-1-2075, AFOSR Grant No.

FA9550-16-1-0139, U.S. Department of Energy Grant No. DE-SC0010064 and DE-SC0014043, and NSF Grant No. 1734315. The work was also supported by the National Natural Science Foundation of China (NSFC) under Grants No. 11991071, 11875175 and 11991073. The authors thank Dr. Mark Palmer, Prof. Vladimir Litvinenko and the ATF staff for their support and hospitality throughout the work done at ATF.

## References

[1] E.S. Weibel, Phys. Rev. Lett. **2**, 83 (1959).

[2] R.L. Morse, Phys. Fluids **14**, 830 (1971).

[3] R.A. Fonseca, L.O. Silva, J.W. Tonge, W.B. Mori, and J.M. Dawson, Physics of Plasmas **10**, 1979 (2003).

[4] B.D. Fried, Phys. Fluids **2**, 337 (1959).

[5] A. Bret, L. Gremillet, and M.E. Dieckmann, Physics of Plasmas **17**, 120501 (2010).

[6] R.M. Kulsrud and S.W. Anderson, ApJ **396**, 606 (1992).

[7] M.V. Medvedev and A. Loeb, ApJ **526**, 697 (1999).

[8] Y. Lyubarsky and D. Eichler, ApJ **647**, 1250 (2006).

[9] K. -I. Nishikawa, P. Hardee, G. Richardson, R. Preece, H. Sol, and G.J. Fishman, ApJ **595**, 555 (2003).

[10] D. Caprioli and A. Spitkovsky, ApJ **765**, L20 (2013).

[11] R. Blandford and D. Eichler, Physics Reports **154**, 1 (1987).

[12] A. Macchi, A. Antonicci, S. Atzeni, D. Batani, F. Califano, F. Cornolti, J.J. Honrubia, T.V. Lisseikina, F. Pegoraro, and M. Temporal, Nucl. Fusion **43**, 362 (2003).

[13] L.O. Silva, R.A. Fonseca, J.W. Tonge, W.B. Mori, and J.M. Dawson, Physics of Plasmas **9**, 2458 (2002).

[14] C. Ren, M. Tzoufras, F.S. Tsung, W.B. Mori, S. Amorini, R.A. Fonseca, L.O. Silva, J.C. Adam, and A. Heron, Phys. Rev. Lett. **93**, 185004 (2004).

[15] T. -Y. B. Yang, J. Arons, and A.B. Langdon, Physics of Plasmas **1**, 3059 (1994).

[16] R.A. Fonseca, L.O. Silva, J. Tonge, R.G. Hemker, J.M. Dawson, and W.B. Mori, IEEE Transactions on Plasma Science **30**, 28 (2002).

[17] P. Arnold, J. Lenaghan, G.D. Moore, and L.G. Yaffe, Phys. Rev. Lett. **94**, 072302 (2005).

[18] C.K. Li, F.H. Séguin, J.A. Frenje, J.R. Rygg, R.D. Petrasso, R.P.J. Town, P.A. Amendt, S.P. Hatchett, O.L. Landen, A.J. Mackinnon, P.K. Patel, V.A. Smalyuk, T.C. Sangster, and J.P. Knauer, Phys. Rev. Lett. **97**, 135003 (2006).

[19] N.L. Kugland, D.D. Ryutov, P.-Y. Chang, R.P. Drake, G. Fiksel, D.H. Froula, S.H. Glenzer, G. Gregori, M. Grosskopf, M. Koenig, Y. Kuramitsu, C. Kuranz, M.C. Levy, E. Liang, J. Meinecke, F. Miniati, T. Morita, A. Pelka, C. Plechaty, R. Presura, A. Ravasio, B.A. Remington, B. Reville, J.S. Ross, Y. Sakawa, A. Spitkovsky, H. Takabe, and H.-S. Park, Nature Physics **8**, 809 (2012).

[20] W. Fox, G. Fiksel, A. Bhattacharjee, P.-Y. Chang, K. Germaschewski, S.X. Hu, and P.M. Nilson, Phys. Rev. Lett. **111**, 225002 (2013).

[21] C.M. Huntington, F. Fiuza, J.S. Ross, A.B. Zylstra, R.P. Drake, D.H. Froula, G. Gregori, N.L. Kugland, C.C. Kuranz, M.C. Levy, C.K. Li, J. Meinecke, T. Morita, R. Petrasso, C. Plechaty,

B.A. Remington, D.D. Ryutov, Y. Sakawa, A. Spitkovsky, H. Takabe, and H.-S. Park, Nature Phys **11**, 173 (2015).

[22] C. Ruyer, S. Bolaños, B. Albertazzi, S.N. Chen, P. Antici, J. Böker, V. Dervieux, L. Lancia, M. Nakatsutsumi, L. Romagnani, R. Shepherd, M. Swantusch, M. Borghesi, O. Willi, H. Pépin, M. Starodubtsev, M. Grech, C. Riconda, L. Gremillet, and J. Fuchs, Nat. Phys. (2020).

[23] G.F. Swadling, C. Bruulsema, F. Fiuza, D.P. Higginson, C.M. Huntington, H.-S. Park, B.B. Pollock, W. Rozmus, H.G. Rinderknecht, J. Katz, A. Birkel, and J.S. Ross, Phys. Rev. Lett. **124**, 215001 (2020).

[24] B. Allen, V. Yakimenko, M. Babzien, M. Fedurin, K. Kusche, and P. Muggli, Phys. Rev. Lett. **109**, 185007 (2012).

[25] C.M. Huntington, A.G.R. Thomas, C. McGuffey, T. Matsuoka, V. Chvykov, G. Kalintchenko, S. Kneip, Z. Najmudin, C. Palmer, V. Yanovsky, A. Maksimchuk, R.P. Drake, T. Katsouleas, and K. Krushelnick, Phys. Rev. Lett. **106**, 105001 (2011).

[26] A. Sampath, X. Davoine, S. Corde, L. Gremillet, M. Gilljohann, M. Sangal, C.H. Keitel, R. Ariniello, J. Cary, H. Ekerfelt, C. Emma, F. Fiuza, H. Fujii, M. Hogan, C. Joshi, A. Knetsch, O. Kononenko, V. Lee, M. Litos, K. Marsh, Z. Nie, B. O'Shea, J.R. Peterson, P.S.M. Claveria, D. Storey, Y. Wu, X. Xu, C. Zhang, and M. Tamburini, Phys. Rev. Lett. **126**, 064801 (2021).

[27] P.S.M. Claveria, X. Davoine, J.R. Peterson, M. Gilljohann, I. Andriyash, R. Ariniello, H. Ekerfelt, C. Emma, J. Faure, S. Gessner, M. Hogan, C. Joshi, C.H. Keitel, A. Knetsch, O. Kononenko, M. Litos, Y. Mankovska, K. Marsh, A. Matheron, Z. Nie, B. O'Shea, D. Storey, N. Vafaei-Najafabadi, Y. Wu, X. Xu, J. Yan, C. Zhang, M. Tamburini, F. Fiuza, L. Gremillet, and S. Corde, ArXiv:2106.11625 [Physics] (2021).

[28] G. Raj, O. Kononenko, M.F. Gilljohann, A. Doche, X. Davoine, C. Caizergues, Y.-Y. Chang, J.P. Couperus Cabadağ, A. Debus, H. Ding, M. Förster, J.-P. Goddet, T. Heinemann, T. Kluge, T. Kurz, R. Pausch, P. Rousseau, P. San Miguel Claveria, S. Schöbel, A. Siciak, K. Steiniger, A. Tafzi, S. Yu, B. Hidding, A. Martinez de la Ossa, A. Irman, S. Karsch, A. Döpp, U. Schramm, L. Gremillet, and S. Corde, Phys. Rev. Research **2**, 023123 (2020).

[29] A. Benedetti, M. Tamburini, and C.H. Keitel, Nature Photon **12**, 319 (2018).

[30] S. Mondal, V. Narayanan, W.J. Ding, A.D. Lad, B. Hao, S. Ahmad, W.M. Wang, Z.M. Sheng, S. Sengupta, P. Kaw, A. Das, and G.R. Kumar, Proceedings of the National Academy of Sciences **109**, 8011 (2012).

[31] G. Chatterjee, K.M. Schoeffler, P. Kumar Singh, A. Adak, A.D. Lad, S. Sengupta, P. Kaw, L.O. Silva, A. Das, and G.R. Kumar, Nat Commun **8**, 15970 (2017).

[32] S. Zhou, Y. Bai, Y. Tian, H. Sun, L. Cao, and J. Liu, Phys. Rev. Lett. **121**, 255002 (2018).

[33] D.V. Romanov, V.Yu. Bychenkov, W. Rozmus, C.E. Capjack, and R. Fedosejevs, Phys. Rev. Lett. **93**, 215004 (2004).

[34] C. Zhang, J. Hua, Y. Wu, Y. Fang, Y. Ma, T. Zhang, S. Liu, B. Peng, Y. He, C.-K. Huang, K.A. Marsh, W.B. Mori, W. Lu, and C. Joshi, Phys. Rev. Lett. **125**, 255001 (2020).

[35] C.-K. Huang, C.-J. Zhang, K.A. Marsh, C.E. Clayton, and C. Joshi, Plasma Phys. Control. Fusion **62**, 024011 (2020).

[36] C. Zhang, C.-K. Huang, K.A. Marsh, C.E. Clayton, W.B. Mori, and C. Joshi, Sci. Adv. **5**, eaax4545 (2019).

[37] R.A. Fonseca, L.O. Silva, F.S. Tsung, V.K. Decyk, W. Lu, C. Ren, W.B. Mori, S. Deng, S. Lee, T. Katsouleas, and J.C. Adam, in *Computational Science — ICCS 2002*, edited by P.M.A. Sloot, A.G. Hoekstra, C.J.K. Tan, and J.J. Dongarra (Springer Berlin Heidelberg, Berlin, Heidelberg, 2002), pp. 342–351.

[38] C.J. Zhang, J.F. Hua, X.L. Xu, F. Li, C.-H. Pai, Y. Wan, Y.P. Wu, Y.Q. Gu, W.B. Mori, C. Joshi, and W. Lu, Sci Rep **6**, 29485 (2016).

[39] N.L. Kugland, D.D. Ryutov, C. Plechaty, J.S. Ross, and H.-S. Park, Review of Scientific Instruments **83**, 101301 (2012).

[40] M.F. Kasim, L. Ceurvorst, N. Ratan, J. Sadler, N. Chen, A. Sävert, R. Trines, R. Bingham, P.N. Burrows, M.C. Kaluza, and P. Norreys, Phys. Rev. E **95**, 023306 (2017).

[41] Q. Du, V. Faber, and M. Gunzburger, SIAM Rev. **41**, 637 (1999).

[42] I.V. Pogorelsky and I. Ben-Zvi, Plasma Phys. Control. Fusion **56**, 084017 (2014).